# A Python Program for Computation of Transition Probabilities, Oscillator Strengths for Li-like ions

[1]M. Hani Zaheer, [1]M. Bilal Khan, [2]S, M, Zeeshan Iqbal, [2]Zaheer Uddin

[1]University of Delaware, Delaware, USA

[2]Department of Physics, University of Karachi, Karachi, Pakistan

**Abstract**

The available software to study the spectroscopic characteristics of atoms, ions, and molecules runs on a server, e.g., the general-purpose atomic structure package (GRASP) and R-matrix method. A Python program has been developed to compute Transition Probabilities, oscillator strengths, Line strengths, matrix elements, and radii of the orbit for lithium and its iso-electronic sequence. The program is straightforward, easily applicable without installation, and uses built-in Python libraries. It can be run on personal computers core I3 and above. The effective charge, effective quantum numbers, and energies of upper and lower levels serve as input parameters for computing the spectral quantities mentioned above. As a case study, we implemented our program on Li I, F VII, Na IX, Al XI, Mg X, and Fe XXIV to calculate transition probabilities, oscillator strengths, and line strengths. The results are compared and found to align with the corresponding values in the NIST data.

**Keywords:** Oscillator strength, Transition probability, Line strength, Lithium-like ions, lithium

Introduction

Advancements in theoretical and applied physics rely heavily on our comprehension of atomic systems. Despite continual progress in experimental techniques for determining atomic properties over the past century, certain atomic states are elusive. Gaps in experimental data also arise due to the prioritization of research interests and funding. In such cases, theoretical predictions of atomic properties are crucial in guiding researchers lacking specific experimental data. Computational techniques, evolving alongside experimental developments, have seen various stages of progress. While state-of-the-art computational methods like coupled cluster, configuration interaction, and density functional theory offer excellent results, they demand significant computational resources beyond the reach of the average researcher. Consequently, there is a growing need for computational methods and user-friendly programs capable of calculating spectral quantities with reasonable accuracy, eliminating the dependency on extraordinary computational resources. Highly-charged ions are an active area of research with applications in quantum metrology and high-resolution spectroscopy [1]. For most experimental and astronomical purposes, sufficiently accurate spectral data tables suffice. NIST has published a database of atomic spectral quantities that can serve as a reference for previous work and the comparison purposes of our results. [2]. Zaidel et al. collected and published tables of spectral lines from many resources for many atoms, including the ones we are investigating [3]. Martin and Wiese published tables of critically evaluated oscillator strengths for lithium isoelectronic sequence [4]. Various experimental investigations into Lithium-like ions have been carried out using different methods. Feldman, Cohen, and Behring used a grazing-incidence spectrometer to observe spectra in lithium-like Magnesium, Aluminum, and Silicon [5]. Fawcett studied plasmas to determine spectral lines in various ions, including Na IX, Mg X, and Al XI [6]. Fawcett used observations of laser-produced plasmas to study various transitions, including some lithium-like transitions [7]. Nahar used the Breit-Pauli R Matrix method to calculate various properties, such as fine structure energy levels and oscillator strengths, of Li I-like atoms, including F VII, Na IX, Al XI, and Mg X [8]. Several experimental and computational studies on single ions in the Lithium isoelectronic sequence have also been carried out. Aggarwal, Keenan, and Heeter performed calculations of energy levels and spectral rates for some Li I-like atoms, including F VII and Na IX, using a general-purpose atomic structure package (GRASP) [9]. Trabert et al. used the beam-foil method to study the fine structure transitions in Mg X [10].

**Theory**

The transition probability (TP), oscillator strength (OS), and line strength (LS) depend on the value of the matrix element given below [33]

$$P^{(1)}_{l_i l_f} = l_> <n_i, l_i|r|n_f, l_f> = l_> \int_0^\infty r^3 R_{n_i l_i} R_{n_f l_f} dr \tag{1}$$

Where $R_{nl}$ is the wave function. The following function works well for lithium atoms and ions of its iso-electronic sequence.

$$R = \left(\frac{2Z^*}{n^*}\right)^{l^*+\frac{3}{2}} \sqrt{\frac{(n^*-l^*-1)!}{2n^*\Gamma(n^*+l^*+1)}}$$
$$\times \exp\left(-\frac{Z^*r}{n^*}\right) r^{l^*} L^{2l^*+1}_{n^*-l^*-1}\left(\frac{2Z^*r}{n^*}\right) \quad (2)$$

The line strength is a function of the matrix elements and is given by [34]

$$S_{LS} = [J_f, J_i, L_f, L_i] \left(\begin{Bmatrix} L_f & S & J_f \\ J_i & 1 & L_i \end{Bmatrix} \begin{Bmatrix} L_f & l_f & L_c \\ 1 & L_i & l_i \end{Bmatrix} P^{(1)}_{l_i l_f}\right)^2 \quad (3)$$

The oscillator strengths and transition probabilities can be calculated with the help of equations (4) and (5).

$$f_{fi} = 3.03672 \times 10^{-6} S_{LS} * \frac{E_f - E_i}{(2*J_i + 1)} \quad (4)$$

$$A_{fi} = 2.0261 \times 10^{-6} \frac{(E_f - E_i)^3}{2J_f + 1} S_{LS} \quad (5)$$

**Computer program**

A user-friendly Python program has been developed to compute matrix elements. Unlike other atomic structure programs, it runs on personal computers with Intel Core I3 and higher. The program is simple and requires characteristics of the upper and lower levels of the transition. The input includes principal quantum numbers, orbital & total angular momenta, and energies of upper and lower levels; it also needs the Ionization limit of the atoms or ions. The built-in functions of associated Laguerre polynomials are used to calculate matrix elements. The Wigner 6 J symbols are used to calculate the line strengths. The program can be downloaded from the Journal website.

**Results and Discussion**

A Python program has been developed to calculate spectral transition parameters, e.g., transition probabilities (TP), oscillator strength (OS), and line strength (OS). The program's performance is checked by calculating and comparing these parameters with the NIST values. We calculated TP, OS, and LS for Li I, F VII, Na IX, Mg X, Al XI, and Fe XXIV ions. The necessary data is obtained from the NIST site, which includes principal quantum numbers, angular momenta, and energies of the Li atom and its iso-electronic sequence. NIST spectral line data contains TP, OS, and LS for various transitions with uncertainties A, AA, AA, +A, B, +B, C, and D. The most reliable transitions are those labeled by A's. We only targeted these transitions, calculated TP, OS, and LS using a Python program, and compared them with the corresponding values. The complete results are shown in supplementary (the results of 10 transitions of each ion are shown here in tables 1 to 6); each contains five columns. The first two columns give the states and angular momenta of upper and lower levels. The third, fourth, and fifth columns give transition probabilities, oscillator strengths, and line strengths. Each of these columns is further divided into three sub-columns. The first columns give the results of this work, the second gives the corresponding values from NIST, if available, and the third gives the percentage error between columns one and two.

**Transitions in Li I**

The transition probabilities, oscillator strengths, and line strengths of $1s^2\,np \to 1s^2\,ms$, $1s^2\,ns \to 1s^2\,mp$, $1s^2\,np \to md$, $1s^2\,nd \to 1s^2\,mp$, $1s^2\,nd \to 1s^2\,mf$ states have been calculated and compared to values in NIST database where A, A+, AA, and AAA label the uncertainties. A total of 195 transitions are compared, of which 184 have less than 5% difference. In contrast, only 11 transitions have percentage errors between 5% and 10%.

**Transitions in F VII**

The transition probabilities, oscillator strengths, and line strengths of the same series as in Li I have been calculated and compared with corresponding values in the NIST database. A total of 363 transitions were considered, of which 278 have a percent error of less than 10%. The remaining 85 transitions have large deviations.

**Transitions in Na IX**

The transition probabilities, oscillator strengths, and line strengths for the same series except for $1s^2\,nd \to 1s^2\,mf$ states have been calculated and compared with the values in the NIST database. Eighty-five transitions were found in the NIST database with lower uncertainties; the transition probabilities, oscillator strengths, and line strengths all have less than 4% percent errors.

**Transitions in Al XI**

A total of 140 transitions from NIST belonging to series $1s^2\,np \to 1s^2\,ms$, $1s^2\,ns \to 1s^2\,mp$, $1s^2\,np \to md$, $1s^2\,nd \to 1s^2\,mp$ were considered, and their transition probabilities, oscillator strengths, and line strengths were calculated and compared. Of these, 130 transitions have uncertainties of less than 5%; the remaining 10 transitions have more than 5% errors. A few of them have large uncertainties.

**Transitions in Mg X**

The NIST database has 79 transitions with lower uncertainties for the series $1s^2\, np \rightarrow 1s^2\, ms$, $1s^2\, ns \rightarrow 1s^2\, mp$, $1s^2\, np \rightarrow md$, $1s^2\, nd \rightarrow 1s^2\, mp$. The calculated transition probabilities, oscillator, and line strengths agree with the NIST values. The percentage of errors in all 79 transitions is less than 5%.

**Transitions in Fe XXIV**

In this work, we also calculated transition probabilities, oscillator strength, and line strength of Fe XXIV. This NIST database has no values with uncertainties labeled by A's. We calculated the TP, OS, and LS for 130 transitions, 58 of which values belong to the transitions in the NIST database. The percent error for 55 transitions is less than 5%; 5 have a percent error between 7-18%, while 72 transitions in XXIV are new.

Table 1: The Transition Probabilities, Oscillator strengths, and Line Strengths for Li I transitions compared with corresponding NIST values.

| State | | Angular momentum | | TP | | | OS | | | LS | | |
|---|---|---|---|---|---|---|---|---|---|---|---|---|
| Lower | Upper | Lower | Upper | This work | NIST | %Diff. | This work | NIST | %Diff. | This work | NIST | %Diff. |
| 2s | 2p | 0.5 | 0.5 | 3.87 E+07 | 3.69E+07 | 4.94 | 0.2608 | 2.49E-01 | 4.75 | 11.5230 | 1.10E+01 | 4.75 |
| 2s | 2p | 0.5 | 1.5 | 3.87 E+07 | 3.69E+07 | 4.94 | 0.5217 | 4.98E-01 | 4.76 | 23.0460 | 2.20E+01 | 4.76 |
| 2p | 3s | 0.5 | 0.5 | 1.10 E+07 | 1.12E+07 | 0.83 | 0.1094 | 1.11E-01 | 1.00 | 5.8546 | 5.91E+00 | 1.01 |
| 2p | 3s | 1.5 | 0.5 | 2.21 E+07 | 2.23E+07 | 0.82 | 0.1094 | 1.11E-01 | 1.00 | 11.7098 | 1.18E+01 | 1.00 |
| 2s | 3p | 0.5 | 0.5 | 1.00 E+06 | 1.00E+06 | 0.26 | 0.0016 | 1.57E-03 | 0.12 | 0.0335 | 3.34E-02 | 0.11 |
| 2s | 3p | 0.5 | 1.5 | 1.00 E+06 | 1.00E+06 | 0.26 | 0.0031 | 3.14E-03 | 0.12 | 0.0669 | 6.69E-02 | 0.11 |
| 2p | 3d | 0.5 | 1.5 | 5.97 E+07 | 5.71E+07 | 4.57 | 0.6666 | 6.39E-01 | 4.38 | 26.7950 | 2.57E+01 | 4.38 |
| 2p | 3d | 1.5 | 1.5 | 1.19 E+07 | 1.14E+07 | 4.57 | 0.0667 | 6.39E-02 | 4.39 | 5.3592 | 5.13E+00 | 4.39 |
| 2p | 3d | 1.5 | 2.5 | 7.17 E+07 | 6.86E+07 | 4.57 | 0.5999 | 5.75E-01 | 4.39 | 48.2324 | 4.62E+01 | 4.39 |
| 3s | 3p | 0.5 | 0.5 | 3.82 E+06 | 3.74E+06 | 2.29 | 0.4137 | 4.05E-01 | 2.12 | 73.2362 | 7.17E+01 | 2.11 |
| 3s | 3p | 0.5 | 1.5 | 3.82 E+06 | 3.74E+06 | 2.34 | 0.8275 | 8.10E-01 | 2.16 | 146.4914 | 1.43E+02 | 2.16 |

Table 2: The Transition capabilities, Oscillator strengths, and Line Strengths for F VII transitions compared with corresponding NIST values.

| State | | Angular momentum | | TP | | | OS | | | LS | | |
|---|---|---|---|---|---|---|---|---|---|---|---|---|
| Lower | Upper | Ji | Jf | This work | NIST | %Diff. | This work | NIST | %Diff. | This work | NIST | %Diff. |
| 2s | 2p | 0.5 | 0.5 | 4.91 E+08 | 4.80E+08 | 2.24 | 0.05827 | 5.71E-02 | 2.05 | 0.34175 | 3.35E-01 | 2.02 |
| 2s | 2p | 0.5 | 1.5 | 5.04 E+08 | 4.93E+08 | 2.26 | 0.11766 | 1.15E-01 | 2.05 | 0.68411 | 6.71E-01 | 1.95 |
| 2p | 3s | 0.5 | 0.5 | 9.69 E+09 | 9.60E+09 | 0.98 | 0.02630 | 2.62E-02 | 0.38 | 0.02332 | 2.33E-02 | 0.07 |
| 2p | 3s | 1.5 | 0.5 | 1.94 E+10 | 1.93E+10 | 0.70 | 0.02643 | 2.63E-02 | 0.51 | 0.04693 | 4.67E-02 | 0.49 |
| 2s | 3p | 0.5 | 0.5 | 4.93 E+10 | 5.05E+10 | 2.36 | 0.09409 | 9.65E-02 | 2.50 | 0.06995 | 7.18E-02 | 2.58 |
| 2s | 3p | 0.5 | 1.5 | 4.91 E+10 | 5.03E+10 | 2.23 | 0.18754 | 1.92E-01 | 2.47 | 0.13938 | 1.43E-01 | 2.53 |

| | | | | | | | | | | | |
|---|---|---|---|---|---|---|---|---|---|---|---|
| 2p | 3d | 0.5 | 1.5 | 1.39E+11 | 1.35E+11 | 3.24 | 0.67710 | 6.58E-01 | 2.90 | 0.56883 | 5.53E-01 | 2.86 |
| 2p | 3d | 1.5 | 1.5 | 2.77 E+10 | 2.69E+10 | 3.14 | 0.06776 | 6.58E-02 | 2.98 | 0.11399 | 1.11E-01 | 2.88 |
| 2p | 3d | 1.5 | 2.5 | 1.66E+11 | 1.61E+11 | 3.28 | 0.60980 | 5.92E-01 | 3.01 | 1.02574 | 9.97E-01 | 2.88 |
| 2p | 4s | 0.5 | 0.5 | 3.73 E+09 | 3.81E+09 | 2.03 | 0.00528 | 5.40E-03 | 2.27 | 0.00338 | 3.46E-03 | 2.39 |

Table 3. The Transition Probabilities, Oscillator strengths, and Line Strengths for Na IX transitions compared with corresponding NIST values.

| State | | Angular momentum | | TP | | | OS | | | LS | |
|---|---|---|---|---|---|---|---|---|---|---|---|
| Lower | Upper | $J_i$ | $J_f$ | This work | NIST | %Diff. | This work | NIST | %Diff. | This work | NIST |
| 2s | 2p | 0.5 | 0.5 | 6.36 E+08 | 6.23 E+08 | 2.1406 | 0.0459 | 0.045 | 1.9521 | 0.2097 | 0.206 |
| 2s | 2p | 0.5 | 1.5 | 6,73 E+08 | 6,60 E+08 | 1.9365 | 0.0936 | 0.0919 | 1.8166 | 0.4200 | 0.413 |
| 2p | 3s | 0.5 | 0.5 | 2,36 E+10 | 2,31 E+10 | 2.3481 | 0.0233 | 0.0228 | 2.0984 | 0.0124 | 0.0122 |
| 2p | 3s | 1.5 | 0.5 | 4.75 E+10 | 4,62 E+10 | 2.8140 | 0.0235 | 0.0229 | 2.5524 | 0.0251 | 0.0245 |
| 2s | 3p | 0.5 | 0.5 | 1.39E+11 | 1.4E+11 | 0.6424 | 0.1037 | 0.105 | 1.2128 | 0.0482 | 0.0486 |
| 3s | 3p | 0.5 | 0.5 | 7,95 E+07 | 7,93 E+07 | 0.2849 | 0.0766 | 0.0765 | 0.1140 | 1.2792 | 1.28 |
| 2s | 3p | 0.5 | 1.5 | 1.39E+11 | 1.4E+11 | 1.0421 | 0.2064 | 0.209 | 1.2457 | 0.0959 | 0.0971 |
| 3s | 3p | 0.5 | 1.5 | 8,43 E+07 | 8,40 E+07 | 0.3516 | 0.1563 | 0.156 | 0.1773 | 2.5608 | 2.56 |
| 2p | 3d | 0.5 | 1.5 | 3.76E+11 | 3.65E+11 | 3.1263 | 0.6802 | 0.662 | 2.7468 | 0.3480 | 0.339 |
| 2p | 3d | 1.5 | 1.5 | 7.50 E+10 | 7,27 E+10 | 3.2557 | 0.0681 | 0.0662 | 2.8730 | 0.0698 | 0.0679 |

Table 4. The Transition Probabilities, Oscillator strengths, and Line Strengths for Mg X transitions and compared with corresponding NIST values.

| State | | Angular momentum | | TP | | | OS | | | LS | | |
|---|---|---|---|---|---|---|---|---|---|---|---|---|
| Lower | Upper | $J_i$ | $J_f$ | This work | NIST | %Diff. | This work | NIST | %Diff. | This work | NIST | %Diff. |

| 2s | 2p | 0.5 | 0.5 | 7.09 E+08 | 6.95E+08 | 2.11 | 0.0415 | 4.07E-02 | 1.89 | 0.1706 | 1.67E-01 | 2.17 |
| 2s | 2p | 0.5 | 1.5 | 7.65 E+08 | 7.51E+08 | 1.91 | 0.0852 | 8.38E-02 | 1.62 | 0.3419 | 3.36E-01 | 1.75 |
| 2p | 3s | 0.5 | 0.5 | 3.45 E+10 | 3.39E+10 | 2.06 | 0.0223 | 2.19E-02 | 1.77 | 0.0096 | 9.49E-03 | 1.46 |
| 2p | 3s | 1.5 | 0.5 | 6.96 E+10 | 6.81E+10 | 2.20 | 0.0225 | 2.21E-02 | 1.97 | 0.0195 | 1.92E-02 | 1.67 |
| 2s | 3p | 0.5 | 0.5 | 2.14E+11 | 2.17E+11 | 1.18 | 0.1074 | 1.09E-01 | 1.45 | 0.0409 | 4.16E-02 | 1.64 |
| 3s | 3p | 0.5 | 0.5 | 8.87 E+07 | 8.85E+07 | 0.31 | 0.0692 | 6.91E-02 | 0.13 | 1.0396 | 1.04E+00 | 0.04 |
| 2s | 3p | 0.5 | 1.5 | 2.13E+11 | 2.16E+11 | 1.27 | 0.2133 | 2.16E-01 | 1.23 | 0.0812 | 8.25E-02 | 1.57 |
| 3s | 3p | 0.5 | 1.5 | 9.71 E+07 | 9.67E+07 | 0.42 | 0.1427 | 1.42E-01 | 0.50 | 2.0819 | 2.08E+00 | 0.09 |
| 2p | 3d | 0.5 | 1.5 | 5.71E+11 | 5.48E+11 | 4.38 | 0.6814 | 6.55E-01 | 4.03 | 0.2830 | 2.72E-01 | 4.06 |
| 2p | 3d | 1.5 | 1.5 | 1.14E+11 | 1.09E+11 | 4.58 | 0.0682 | 6.56E-02 | 4.02 | 0.0568 | 5.47E-02 | 3.90 |

Table 5: The Transition Probabilities, Oscillator strengths, and Line Strengths for Al XI transitions compared with corresponding NIST values.

| State | | Angular momentum | | TP+A1:O36 | | | OS | | | LS | | |
|---|---|---|---|---|---|---|---|---|---|---|---|---|
| Lower | Upper | Lower | Upper | This work | NIST | %Diff. | This work | NIST | %Diff. | This work | NIST | %Diff. |
| 7p | 8d | 0.5 | 1.5 | 8.15E+09 | 9.11E+08 | 10.52 | 0.5601 | 6.27E-01 | 10.67 | 5.5874 | 6.26E+00 | 10.74 |
| 7p | 8d | 1.5 | 1.5 | 1.63 E+09 | 1.82E+08 | 10.42 | 0.0560 | 6.27E-02 | 10.67 | 1.1175 | 1.25E+00 | 10.60 |
| 7p | 8d | 1.5 | 2.5 | 9.78 E+09 | 1.09E+09 | 10.25 | 0.5041 | 5.64E-01 | 10.62 | 10.0574 | 1.13E+01 | 11.00 |
| 6p | 8d | 0.5 | 1.5 | 1.36 E+09 | 1.42E+09 | 3.71 | 0.1512 | 1.57E-01 | 3.69 | 0.6052 | 6.28E-01 | 3.64 |
| 6p | 8d | 1.5 | 2.5 | 1.64 E+09 | 1.70E+09 | 3.49 | 0.1361 | 1.41E-01 | 3.49 | 1.0893 | 1.13E+00 | 3.60 |
| 6p | 8d | 1.5 | 1.5 | 2.73 E+08 | 2.83E+08 | 3.37 | 0.0151 | 1.57E-02 | 3.69 | 0.1210 | 1.26E-01 | 3.94 |
| 4s | 5p | 0.5 | 0.5 | 9.06 E+09 | 9.37E+09 | 3.34 | 0.1291 | 1.34E-01 | 3.68 | 0.2622 | 2.72E-01 | 3.59 |
| 6p | 7d | 0.5 | 1.5 | 1.94 E+09 | 2.01E+09 | 3.24 | 0.5740 | 5.94E-01 | 3.37 | 3.7527 | 3.88E+00 | 3.28 |
| 6p | 7d | 1.5 | 1.5 | 3.89 E+08 | 4.02E+08 | 3.24 | 0.0574 | 5.94E-02 | 3.37 | 0.7505 | 7.77E-01 | 3.41 |
| 6p | 7d | 1.5 | 2.5 | 2.33 E+09 | 2.41E+09 | 3.16 | 0.5166 | 5.35E-01 | 3.44 | 6.7548 | 7.00E+00 | 3.50 |
| 3d | 5p | 1.5 | 0.5 | 2.58 E+09 | 2.66E+09 | 2.82 | 0.0022 | 2.25E-03 | 3.15 | 0.0030 | 3.15E-03 | 3.32 |

Table 6: The Transition Probabilities, Oscillator strengths, and Line Strengths for Fe XXIV transitions compared with corresponding NIST values.

| State | | Angular momentum | | TP | | | OS | | | LS | | |
|---|---|---|---|---|---|---|---|---|---|---|---|---|
| Lower | Upper | Ji | Jf | This work | NIST | %Diff. | This work | NIST | %Diff. | This work | NIST | %Diff. |
| 3p | 5d | 0.5 | 1.5 | 9.73286E+11 | 9.73E+11 | 0.0294 | 0.1379 | 0.139 | 0.7819 | 0.0198 | 0.02 | 1.2126 |
| 2s | 6p | 0.5 | 0.5 | 1.01698E+12 | 1.02E+12 | 0.2965 | 0.0069 | 0.00705 | 2.4161 | 0.0003 | 0.000315 | 3.3186 |
| 4p | 7d | 0.5 | 1.5 | 1.50513E+11 | 1.51E+11 | 0.3224 | 0.0616 | 0.0621 | 0.8795 | 0.0150 | 0.0151 | 0.7886 |
| 2s | 4p | 0.5 | 0.5 | 3.38735E+12 | 3.4E+12 | 0.3720 | 0.0319 | 0.033 | 3.4186 | 0.0017 | 0.0017 | 2.1208 |
| 2p | 6d | 0.5 | 1.5 | 1.52751E+12 | 1.52E+12 | 0.4942 | 0.0218 | 0.0222 | 1.7548 | 0.0010 | 0.00102 | 2.7595 |
| 2p | 5d | 0.5 | 1.5 | 2.82645E+12 | 2.8E+12 | 0.9445 | 0.0451 | 0.046 | 1.8641 | 0.0022 | 0.0022 | 1.3087 |
| 2s | 6p | 0.5 | 1.5 | 1.00904E+12 | 1.02E+12 | 1.0741 | 0.0136 | 0.0141 | 3.2395 | 0.0006 | 0.000629 | 4.0129 |
| 4s | 5p | 0.5 | 1.5 | 2.30308E+11 | 2.33E+11 | 1.1553 | 0.3111 | 0.3159 | 1.5284 | 0.1376 | 0.1398 | 1.5826 |
| 2p | 5d | 1.5 | 1.5 | 5.46279E+11 | 5.4E+11 | 1.1629 | 0.0044 | 0.0045 | 1.1734 | 0.0004 | 0.00044 | 1.8347 |
| 4s | 6p | 0.5 | 0.5 | 1.4416E+11 | 1.46E+11 | 1.2600 | 0.0419 | 0.04261 | 1.7256 | 0.0121 | 0.01238 | 1.8833 |

# Conclusion

A Python program that calculates the transition probabilities, oscillator strengths, and line strengths for lithium and lithium-like ions is developed. Nine hundred and ninety-two spectral lines were studied from Li I, F VII, Na IX, Al Xi, Mg X, and Fe XXIV transitions to check the reliability of the calculated values. The transition probabilities, oscillator strengths, and line strengths were calculated by the Python program and compared with values from the NIST site. Out of 920 transitions, 811 agree well with the NIST values, demonstrating the accuracy of the results. The remaining 109 transitions need to be verified experimentally. TP, OS, and LS of seventy-two transition of Fe XXIV are new. We intend to continue the same calculations for other lithium-like ions whose data is unavailable on the NIST site. This will be useful in astrophysics for analyzing spectra of highly ionized atoms.